\documentclass[11pt,a4paper]{article}
\usepackage{jcappub}
\usepackage[sort&compress]{natbib}

\begin{document}

\title{Strongly lensed gravitational waves from intrinsically faint double compact binaries --- prediction for the Einstein Telescope}

\author{Xuheng Ding,$^1$}
\author{Marek Biesiada,$^{1,2,3}$}
\author{Zong-Hong Zhu $^1$}

\affiliation{
$^1$ Department of Astronomy, Beijing Normal
University, Beijing 100875, China \\
$^2$ Department of Astrophysics and Cosmology, Institute of Physics,
University of Silesia, Uniwersytecka 4, 40-007 Katowice, Poland \\
$^3$ Kavli Institute for Theoretical Physics China, CAS, Beijing 100190, China

}

\abstract{With a fantastic sensitivity improving significantly over the advanced GW detectors,
Einstein Telescope (ET) will be able to observe hundreds of thousand inspiralling
double compact objects per year. By virtue of gravitational lensing effect, intrinsically unobservable faint sources
can be observed by ET due to the magnification by intervening galaxies. We explore the
possibility of observing such faint sources amplified by strong gravitational lensing. Following our previous work, 
we use the merger rates of DCO (NS-NS,BH-NS,BH-BH systems) as calculated by Dominik et al.(2013). 
It turns out that tens to hundreds of such (lensed) extra
events will be registered by ET. This will strongly broaden the ET's distance reach for signals from such coalescences to the redshift range $z = 2 - 8$.
However, with respect to the full inspiral event catalog this magnification bias is at the level of 0.001 and should not affect much cosmological inferences.
 }

\maketitle

\section{Introduction} \label{sec:intro}

Gravitational waves (GW thereafter) are expected to be registered soon by the upgraded advanced LIGO/VIRGO detectors \citep{LIGO,VIRGO}. This will open an era of experimental gravitational wave astrophysics with the expected detection rates of order ranging from tens to a thousand events per year. The real breakthrough will come with the new generation of detectors, an example of which --- the Einstein Telescope (ET) has already went through the initial design study \citep{Abernathy03}.
Because such an instrument will improve an order of magnitude in sensitivity over the advanced laser interferometer detectors LIGO and VIRGO, simple scaling arguments lead us to expect tens to hundreds thousands of detections per year. With such a number of detectable events and with the detector horizon reaching 1 -- 2 $Gpc$ one may expect that non-negligible number of sources would be gravitationally lensed by galaxies lying in between. We considered this problem in our previous papers \citep{JCAP_ET1, JCAP_ET2}. In particular the analysis performed in \citep{JCAP_ET2} was very comprehensive in the sense of taking into account full population of double compact objects (DCO) i.e. NS-NS, NS-BH, and BH-BH binaries, as well as taking the cosmological merger rates at different redshifts as suggested by the population synthesis model (using {\tt StarTrack} code) in \citet{BelczynskiII}. The result was that ET would register about 50 -- 100 strongly lensed inspiral events per year with statistics dominated by the BH-BH systems. These results suggest that ET will provide a considerable catalog of strongly lensed events during a few years of its successful operation.
Our previous estimates in \citep{JCAP_ET2} were obtained under assumption that DCO systems intrinsically have signal to noise ratio (SNR) greater or equal to the threshold $\rho_0 = 8.$ This was a reasonable assumption since we wanted to estimate the rates of lensed events in the population of sources detectable to the ET.
However, gravitational lensing effect will magnify the amplitudes of the GW sources increasing this way the SNR of each lensed DCO system. Therefore, intrinsically faint sources (having $SNR<8$ ) may now become observable to the ET.
In this paper, we supplement our previous study by considering sources which are intrinsically below the detection threshold and consequently would not be observed had not they been magnified by the lens.

The paper is organized as follows. In Section~\ref{sec:methodology}, we briefly recapitulate our methodology (referring to \citep{JCAP_ET1, JCAP_ET2} for detailed calculations) and review DCO catalog build from evolutionary population synthesis code which we use thereafter.
Then, in Section~\ref{sec:conclusions} we present our results and conclusions.

Throughout the paper we comply with the notation and nomenclature of \citep{JCAP_ET2} and \citep{BelczynskiII}. For the sake of consistency with previous works, we assume flat FRW cosmological model. In particular the expansion rate in this model reads:
\begin{equation} \label{H}
H(z) = H_0 \sqrt{\Omega_m (1+z)^3 + (1 - \Omega_m) }
\end{equation}
with $H_0 = 70\;km/s/Mpc$, $\Omega_m = 0.3$ as in \citet{BelczynskiII}. We will also adopt the notation: $E(z)=H(z)/H_0$ and ${\tilde r} = \int_0^{z} \frac{dz'}{E(z')}$ --- the non-dimensional comoving distance.


\section{Methodology} \label{sec:methodology}

Our goal is to estimate the number of GW sources which would be magnified above the detector's threshold. However, unlike in our previous estimates we will admit them being intrinsically faint, i.e. such that they would not be detected haven't they been lensed. From the physical point of view a relevant observable quantity is the dimensionless amplitude of the GW - the strain $h(t) = F_{+} h_{+}(t) + F_{\times} h_{\times}(t)$, where $+$ and $\times$ denote two independent polarizations. However, the detector is not able to register pure signal, but rather the signal is buried in the detector's noise, thus we have to use matched filtering technique in which signal to noise ratio (SNR) $\rho$ is the integrated signal spectral power weighted down by noise power spectral density. In particular, for a single detector it reads
(for more details see \citep{JCAP_ET1} and references therein):

 \begin{equation} \label{SNR}
 \rho = 8 \Theta \frac{r_0}{d_L(z_s)} \left(\frac{{\cal M}_z}{1.2\;M_{\odot}}\right)^{5/6}
 \sqrt{\zeta(f_{max})}
 \end{equation}
 where: $d_L$ is the luminosity distance to the source, $\Theta$ is the orientation factor
 capturing part of sensitivity pattern due to (usually non-optimal) random
 relative orientation of a DCO system with respect to the detector, $r_0$ above is detector's characteristic distance
 parameter. After \citet{TaylorGair} we consider two options:
 the ET initial design, which gives $r_0 = 1527 \; Mpc$ and the advanced ``xylophone'' configuration, which
 gives $r_0 = 1918 \; Mpc$. We also assume that $\zeta(f_{max})=1$ (for justification and more details see \citep{TaylorGair}).
 The orientation factor $\Theta$ is defined as
 \begin{equation} \label{Theta}
 \Theta = 2 [ F_{+}^2(1 + \cos^2{\iota} )^2 + 4 F_{\times}^2 \cos^2{\iota} ]^{1/2}
 \end{equation}
where $F_{+} = \frac{1}{2} (1 + \cos^2{\theta}) \cos{2\phi} \cos{2 \psi} - \cos{\theta} \sin{2 \phi} \sin{ 2 \psi}$ and
$F_{\times} = \frac{1}{2} (1 + \cos^2{\theta}) \sin{2\phi} \cos{2 \psi} + \cos{\theta} \sin{2 \phi} \cos{ 2 \psi}$ are the interferometer strain responses to different polarizations of gravitational wave.

Probability distribution for $\Theta$ calculated under assumption of uncorrelated orientation angles $(\theta, \phi, \psi, \iota)$ is known to be of the following form:
\begin{eqnarray} \label{P_theta}
P_{\Theta}(\Theta) &=& 5 \Theta (4 - \Theta)^3 /256, \qquad {\rm if}\;\;\;
0< \Theta < 4  \\
P_{\Theta}(\Theta) &=& 0, \qquad {\rm otherwise} \nonumber
\end{eqnarray}

The yearly detection rate of DCO sources originating at redshift $z_s$ and producing the signal with SNR exceeding the detector's threshold $\rho_0$ (previously assumed that sources intrinsically have SNR parameter $\rho = 8$ but now we relax this assumption) can be expressed as:
 \begin{equation} \label{Ndot}
{\dot N}(>\rho_0|z_s) = \int_0^{z_s} \frac{d {\dot N}(>\rho_0)}{dz} dz
\end{equation}
where
\begin{equation} \label{rate_nl}
\frac{d {\dot N}(>\rho_0)}{dz_s} = 4\pi \left( \frac{c}{H_0} \right)^3 \frac{{\dot n_0}(z_s)}{1+z_s}\;  \frac{{\tilde r}^2(z_s)}{E(z_s)} \; C_{\Theta}(x(z_s, \rho_0))
\end{equation}
is the rate at which we observe the inspiral DCO events (sources) that originate in the redshift interval $[z, z+dz]$. In Eq.(\ref{rate_nl}) ${\dot n_0}(z_s)$ denotes intrinsic coalescence rate in the local Universe at redshift $z_s$,
$C_{\Theta}(x) = \int_x^{\infty} P_{\Theta}(\Theta) d\Theta$ and $x(z, \rho) = \frac{\rho}{8} (1+z)^{1/6} \frac{c}{H_0} \frac{{\tilde r}(z)}{r_0} \left( \frac{1.2\; M_{\odot}}{{\cal M}_0} \right)^{5/6}$

Here, like in the previous paper \citep{JCAP_ET2}, we use the values of inspiral rates ${\dot n_0}(z_s)$ reported by \citep{BelczynskiII} for each redshift slice they considered. To be more specific they evolved binary systems from ZAMS until the compact binary formation under certain well motivated assumptions about star formation rate, galaxy mass distribution, stellar populations, their metallicities and galaxy metallicity evolution with redshift (``low end'' and ``high end'' cases). In order to make straightforward comparison with the results presented in \citep{JCAP_ET2} we
consider all evolutionary scenarios leading to the DCO formation, presented in \citet{BelczynskiII} that is: the standard scenario, optimistic common envelope scenario, delayed SN explosion and high BH kick scenario. 
We have taken the data from the website {\tt http:www.syntheticuniverse.org}, more specifically the so called ``rest frame rates'' in cosmological scenario. For the chirp masses we have assumed the following values: $1.2\;M_{\odot}$ for NS-NS, $3.2\;M_{\odot}$ for BH-NS and $6.7\;M_{\odot}$ for BH-BH systems.
According to \citet{BelczynskiI}, they represent average chirp mass for each category of DCO simulated by population synthesis.

Since we relax the fixed value of the intrinsic signal to noise ratio, instead of Eq.~(\ref{rate_nl}) we have to start with the differential inspiral rate per redshift and per SNR parameter $\rho$:
\begin{equation} \label{diff_rate}
\frac{\partial^2 {\dot N}}{\partial z_s \partial \rho} = 4 \pi \left( \frac{c}{H_0} \right)^3 \frac{{\dot n_0}(z_s)}{1+z_s} \frac{{\tilde r}^2(z_s)}{E(z_s)} P_{\Theta} (x(z_s, \rho)) \frac{x(z_s, \rho)}{\rho}
\end{equation}

Concerning gravitational lensing we adopt the same approach as in our previous paper \citep{JCAP_ET2}, i.e. we assume conservatively that the population of lenses comprise only elliptical galaxies. Therefore, we will model the lenses as singular isothermal spheres (SIS) which is a good approximation of early type galaxies \citep{Koopmans09}.
Characteristic angular scale of lensing phenomenon is set by the Einstein radius, which for the SIS model reads:  $\theta_E = 4 \pi \left( \frac{\sigma}{c} \right)^2 \frac{d_A(z_l,z_s)}{d_A(z_s)} $, where $\sigma$ is the velocity dispersion of stars in lensing galaxy, $d_A(z_l,z_s)$ and $d_A(z_s)$ are angular diameter distances between the lens and the source and to the source, respectively. It is convenient to use the Einstein radius as a unit and convert the angular distance of the image (w.r.t. the center of the lens) $\theta$ or the angular position of the source $\beta$ to respective dimensionless parameters: $x = \frac{\theta}{\theta_E}$, $ y = \frac{\beta}{\theta_E}$. Then the necessary condition for strong lensing (multiple images) is $y<1$. Images (brighter $I_{+}$ and fainter one $I_{-}$ ) form at locations $x_{\pm} = 1 \pm y$ with magnifications: $\mu_{\pm} = \frac{1}{y} \pm 1$.
Hence, the gravitationally lensed GW signal would come from these two images with appropriate relative time delay (see \citep{JCAP_ET1, JCAP_ET2}) and with different amplitudes: $
h_{\pm} = \sqrt{\mu_{\pm}} \; h(t) = \sqrt{\frac{1}{y} \pm 1}\; h(t)
$
where $h(t)$ denotes the intrinsic amplitude (i.e. the one which would have been observed without lensing). Analogous relations are valid for the SNR parameter $\rho$. Assuming the threshold SNR for detection $\rho_0 = 8$, one can observe lensed images (of the source with an intrinsic SNR equal to $\rho_{intr.}$) if the misalignment of the source with respect to the optical axis of the lens satisfies:
\begin{equation} \label{y_condition}
y_{\pm} \leq y_{\pm,max} = \left[ \left( \frac{8}{\rho_{intr.}} \right)^2 \mp 1 \right]^{-1}
\end{equation}

These conditions (for the $I_{+}$ and $I_{-}$ image) influence elementary cross section for lensing (see e.g.\citep{JCAP_ET1}):
\begin{equation} \label{cross_section}
S_{cr, \pm}(\sigma, z_l, z_s, \rho) = \pi \theta_E^2 y_{\pm,max}^2  = 16 \pi^3 \left( \frac{\sigma}{c} \right)^4 \left( \frac{{\tilde r}_{ls}} {{\tilde r}_{s}} \right)^2 y_{\pm, max}^2
\end{equation}
which is necessary to calculate optical depth for lensing leading to magnifications of $I_{+}$ and $I_{-}$ images above the threshold:
\begin{equation} \label{tau}
\tau_{\pm}(z_s, \rho) = \frac{1}{4 \pi} \int_0^{z_s}\; dz_l \; \int^{\infty}_0 \; d \sigma \; 
4 \pi \left( \frac{c}{H_0} \right)^3 \frac{ {\tilde r}_l^2}{E(z_l)} S_{cr, \pm}(\sigma, z_l, z_s) \frac{d n}{d \sigma}
\end{equation}

In analogy to and in order to comply with our previous papers, \citep{JCAP_ET1} and \citep{JCAP_ET2}, we model the velocity dispersion distribution in the population of lensing galaxies as a modified Schechter function $\frac{d n}{d \sigma} = n_{*} \left( \frac{\sigma}{\sigma_{*}} \right)^{\alpha} \exp{\left( - \left( \frac{\sigma}{\sigma_{*}} \right)^{\beta} \right)} \frac{\beta}{\Gamma (\frac{\alpha}{\beta}) } \frac{1}{\sigma}$ with the parameters $n_{*}$,$\sigma_{*}$,${\alpha}$ and $\beta$ we taken after \citet{Choi2007} (for discussion about such choice in view of other data on velocity dispersion distribution functions see \citep{JCAP_ET2}). Optical depth for lensing depends on the survey duration. However, it turns out (e.g. our previous papers) that the survey time (i.e. 1 year, 5 year and continuous) has little effect on the detection rate. Thus, in this work we only consider the case of continuous search.

We can now combine Eq.(\ref{diff_rate}) and Eq.(\ref{tau}) in order to calculate quantities like the lensing rate of intrinsically faint ($\rho < \rho_0$) DCOs having $I_{+}$ or $I_{-}$ images magnified above the threshold $\rho_0$:
\begin{equation} \label{lensing_rate}
{\dot N}_{lensed, \pm} = \int_0^{z_{max}} dz_s \int_0^{\rho_0} \tau_{\pm}(z_s, \rho) \frac{\partial^2 {\dot N}}{\partial z_s \partial \rho} d \rho
\end{equation}
or differential lensing rates with respect to $\rho$ or $z_s$ respectively:
\begin{equation} \label{diff_lensing_rate_rho}
\frac{d {\dot N}_{lensed, \pm}}{d \rho} = \int_0^{z_{max}} \tau_{\pm}(z_s, \rho) \frac{\partial^2 {\dot N}}{\partial z_s \partial \rho} dz_s
\end{equation}
\begin{equation} \label{diff_lensing_rate_zs}
\frac{d {\dot N}_{lensed, \pm}}{d z_s} = \int_0^{\rho_0}  \tau_{\pm}(z_s, \rho) \frac{\partial^2 {\dot N}}{\partial z_s \partial \rho} d \rho
\end{equation}


\begin{table*}[ht]
\footnotesize 
\caption{Expected numbers of lensed GW events with $\rho_{intr}<8$ for which the $I_-$ image is magnified above threshold $\rho_0 = 8$. We also assumed the continuous survey, which in practice means that its duration is longer than 5 years. Nomenclature of DCO formation scenarios and galaxy metallicity evolution follows that of \citet{BelczynskiII}. Predictions for two configurations of the ET are given.
}
\label{lensing-1}
\begin{center}  
\begin{tabular}{cccccc}

\hline
DCO scenario  &standard&optimistic CE&delayed SN& high BH kicks  \\
metallicity evolution &{High; Low}&{High; Low}
&{High; Low}&{High; Low}\\
\hline\\

NS-NS & & & & \\
initial design &0.6; 0.4&5.1; 5.3 &0.6; 0.5&0.6; 0.4\\
xylophone &0.9; 0.8 &9.4; 9.6 & 1.0; 0.9 & 1.0; 0.8\\
\\
\hline\\

BH-NS & & & & \\
initial design &1.2; 1.4&2.2; 2.1 &0.6; 0.7&0.1; 0.2\\
xylophone &1.4; 1.5 &2.4; 2.2 & 0.7; 0.7 & 0.2; 0.2\\
\\
\hline\\

BH-BH & & & & \\
initial design &29.9; 34.1&69.0; 69.5 &25.6; 29.5&2.3; 2.8\\
xylophone &26.6; 29.9 &58.8; 58.6 & 22.9; 26.0 & 2.1; 2.5\\
\\
\hline\\

TOTAL & & & & \\
initial design &31.7;35.9&76.3;77.0&26.8;30.6&3.0;3.4\\
xylophone &29.0;32.2&70.6;70.4&24.6;27.6&3.3;3.5\\
\\
\hline\\

\end{tabular}\\
\end{center}
\end{table*}

\begin{figure}
\begin{center}
\includegraphics[width=70mm]{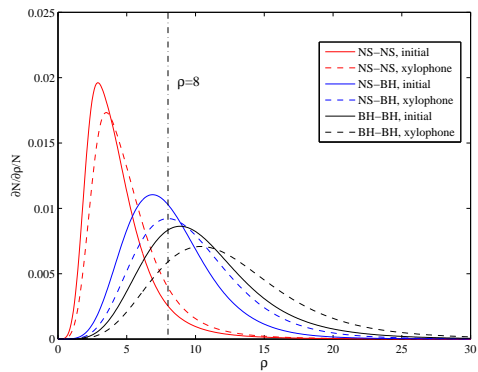}
\end{center}
\caption{The observed lensed GW event number distribution as a function of $\rho$.  ``Low-end'' metallicity galaxy evolution and standard model of DCO formation are assumed.
\label{fig1}}
\end{figure}

\begin{figure*}
\begin{center}
\includegraphics[angle=0,width=70mm]{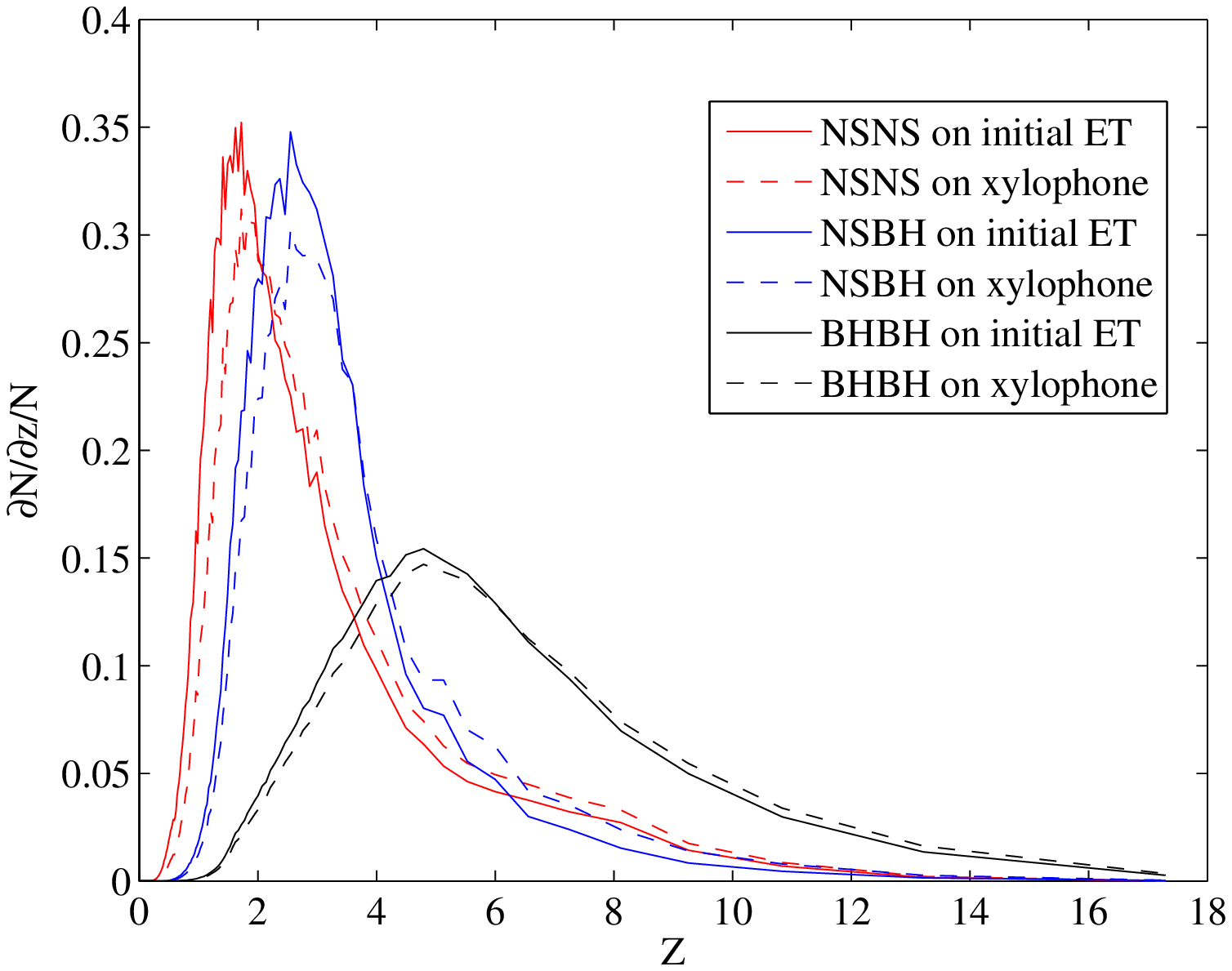}
\includegraphics[angle=0,width=70mm]{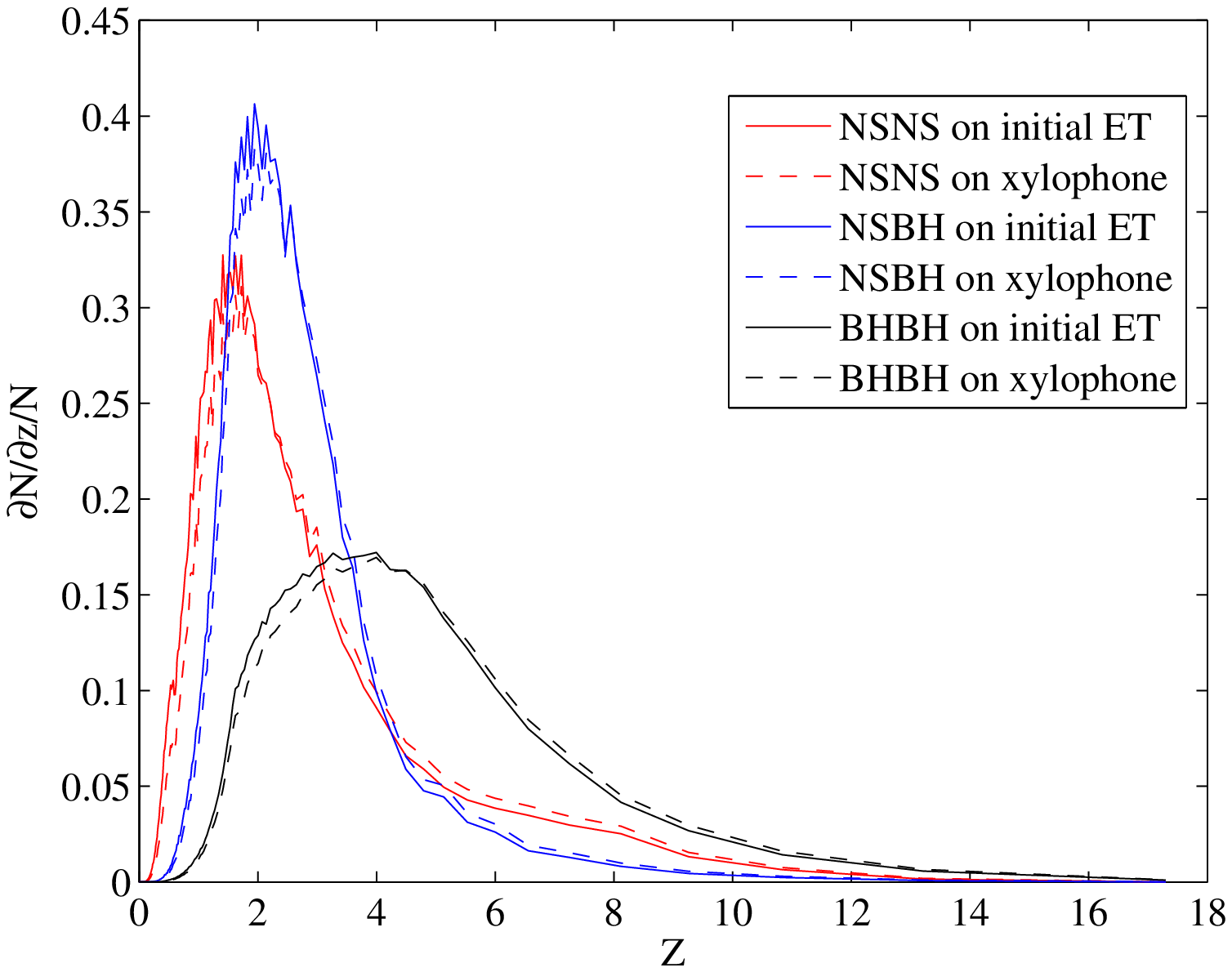}
\caption{The observed lensed GW event number distribution as a function of $z$. Left figure corresponds to $\rho_{intr}<8$ for the $I_-$ image. The right one corresponds to the $I_-$ image including both $\rho_{int}<8$ and $\rho_{int}>8$. ``Low-end'' metallicity galaxy evolution and standard model of DCO formation are assumed.}
\label{fig2}
\end{center}
\end{figure*}

\begin{table*}[ht]
\footnotesize 
\caption{Expected numbers of lensed GW events with $\rho_{intr}<8$ for which the $I_+$ image is magnified above threshold $\rho_0 = 8$. Other assumptions and terminology -- like in Table~\ref{lensing-1}.
}
\label{lensing-2}
\begin{center}  
\begin{tabular}{cccccc}

\hline
DCO scenario  &standard&optimistic CE&delayed SN& high BH kicks  \\
metallicity evolution &{High; Low}&{High; Low}
&{High; Low}&{High; Low}\\
\hline\\

NS-NS & & & & \\
initial design &2.1; 1.5&17.9; 20.4 &2.4; 1.6&2.3; 1.5\\
xylophone &3.9; 3.2 &40.5; 43.5 & 4.4; 3.5 & 4.1; 3.2\\
\\
\hline\\

BH-NS & & & & \\
initial design &6.4;7.1&11.8; 11.3 &3.3; 3.6 &0.7; 0.8\\
xylophone &7.5; 7.9 &12.8; 11.9 & 3.7; 3.8 & 0.9; 0.9\\
\\
\hline\\

BH-BH & & & & \\
initial design &161.8; 184.1&373.4; 376.2 &138.3; 159.3&12.5; 14.9\\
xylophone &144.2; 161.9 &318.5; 317.4 & 124.0; 140.7 & 11.5; 13.6\\
\\
\hline\\

TOTAL & & & & \\
initial design &170.4;192.7&403.0;407.9&144.0;164.5&15.5;17.2\\
xylophone &155.6;173.0&371.8;372.9&132.0;148.0&16.5;17.7\\
\\
\hline\\
\end{tabular}\\
\end{center}
\end{table*}

\begin{figure*}
\begin{center}
\includegraphics[width=70mm]{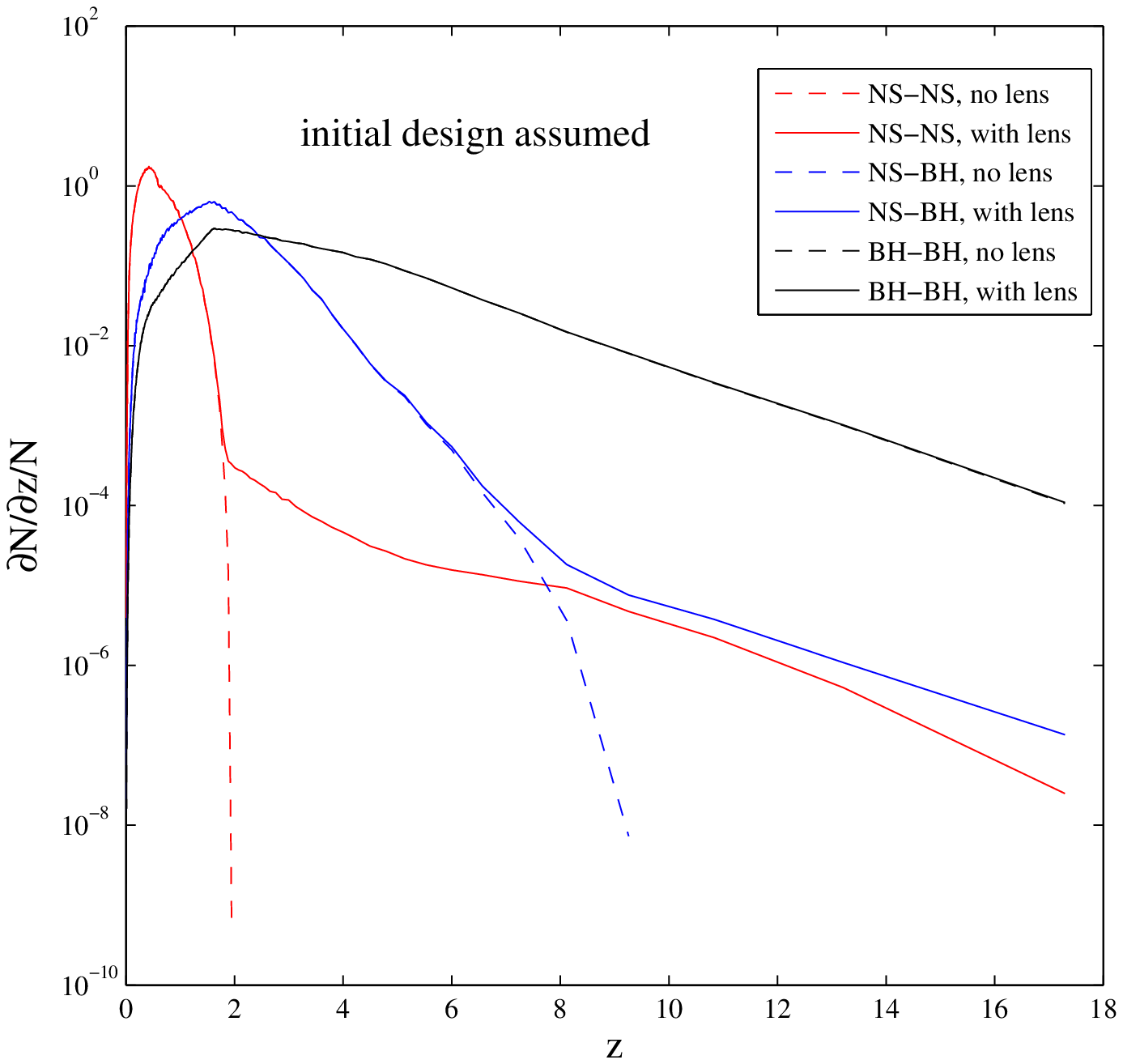}
\caption{Probability density of DCO inspiral events yearly rate (as a function of redshift) to be observed by the ET in its initial design.
Continuous lines refer to the total catalogue of lensed and non-lensed systems, non-lensed systems are represented by dashed lines. It illustrates the magnification bias for different DCO systems. Note the logarithmic scale used in this picture. In particular figure shows that magnification bias is negligibly small, and even not noticeable for BH-BH systems. }

\label{fig3}
\end{center}
\end{figure*}

\section{Results and discussion} \label{sec:results}

Table~\ref{lensing-1} shows the expected yearly detection rate of lensed DCO inspiral events having intrinsic SNR below the threshold of detection $\rho_{intr.} < \rho_0 =8$ and magnified strongly enough for the $I_{-}$ image to be detected. This guarantees that the $I_{+}$ image will be observed, too. It means that in principle one would be able to establish the nature of such two time-delayed signals with similar temporal structure (frequency drift) but differing only in amplitudes, as a consequence of gravitational lensing. Hence, the rates shown in Table~\ref{lensing-1} supplement the rates reported in Table 3 and Table 4 of \citet{JCAP_ET2}. Concerning the total expected rates of lensed GW events, one can see that they are roughly doubled for the initial design of ET, for the ``xylophone'' design this effect is smaller (about $50 - 60 \%$ increase) but still substantial. Therefore the numbers displayed in Table~\ref{lensing-1} quantify magnification bias in the catalogue of lensed GW inspiral events. It is most pronounced for NS-NS systems, where one has an order of magnitude increase in the expected lensing rates.

Figure~\ref{fig1} shows the distribution of yearly rates of lensed events with respect to the intrinsic SNR parameter $\rho$ in different classes of DCO within the reach of ET detector in its initial and ``xylophone'' design. One can see that NS-NS population peaks below $\rho_{intr.}=5$ with a small high-end tail of the distribution being intrinsically stronger than the detection threshold $\rho_0=8$. This is reflected in marginal contribution of NS-NS systems to the lensing rates discussed in \citep{JCAP_ET2} and explains why these systems are affected the most by magnification from strong lensing. However, even for the BH-BH binary systems a noticeable part of the distribution lies below the threshold. Figure~\ref{fig2} displays the distribution of yearly rates of lensed events with respect to the source redshift $z_s$. Left panel shows the systems with intrinsic SNR below the threshold, while the right one shows the combined distribution of the total available DCO population (having SNR below as well as above the detection threshold). One can see that gravitationally lensed intrinsically faint sources probe higher redshifts, so the future catalog of gravitationally lensed GW events would be contaminated by higher redshift sources --- in agreement with the general idea of how magnification bias works. This broadens the ET's distance reach and may enable us listen to the waves of DCO inspirals from redshifts z = 2 -- 8 which will lead to better understanding of the early epochs of star formation.

One of the most important issues in modern observtional cosmology is the determination of
star-formation history. Over last two decades a coherent picture emerged \citep{Heavens} according to which star formation rate (SFR) increase
to the redshift z=2 ($SFR(z=2) \approx 10 \times SFR(z=0)$) and then decreases.
This was possible to achieve because of massive spectroscopic surveys like SDSS and invaluable information the spectrum of a galaxy
bears concerning its star formation history and gas metallicity.
However the behavior of SFR at higher redshifts is still uncertain.
It has been know that coalescence rates probed by GW laser interferometric observations
could be used to test alternative scenarios of star formation. For example the ET Design Study document demonstrates
that the ET will be able to discriminate between four SFR models: \citet{Hopkins}, \citet{Nagamine}, \citet{Fardal} and \citet{Wilkins}.
The DCO coalescence detection rate would be the best source of such constraints, lensed events of intrinsically faint sources
will not add much to it. However, still remains the possibility to measure the luminosity distance and redshifted chirp mass of such sources.
Assuming some reliable background cosmology one would be able to estimate the intrinsic chirp mass of the system.
The possibility to determine DCO masses up to z=5 (or even higher for BH-BH systems) is intriguing.
It is because such distant DCO systems are fossils of the era of high-mass star formation in the Universe.
Let us also remind that detection of coalescing DCO at certain redshift yields information about BH and NS
formed at even earlier epochs because of the delays between formation and coalescence.
Measuring the masses of DCO containing BH will open a new chapter of astrophysics and certainly will tell us a lot about star formation scenarios.
The comprehensive insight into the underlying distribution of NS masses would
provide the means to study not only specific particular aspects like NS matter equation of state but also more fundamental ones.
For example as the masses of NSs also retain information about the past value of the
effective gravitational constant G, with the determination of the NS mass range at high redshifts it may be even possible to
probe the potential evolution of such physical constants \citep{Thorsett}. Equally intriguing question is about the influence of dark matter accumulating in the cores of the NS on their properties (masses, conversion to quark stars etc.). This issue has usually been neglected in astrophysical studies, but recently  attracted growing attention \citep{Bertoni, Silk, Bramante}. In particular \citet{Kouvaris} try to explain the NS braking index by dark matter whereas \citet{Fuller} invoke dark matter induced collapse of NS as a link between fast radio bursts and missing pulsar problem.

In addition to the magnification bias on the lensed events, one can estimate the magnification bias at the level of full DCO inspiral events catalogue. For this purpose one should calculate the detection rate of intrinsically faint events for which only $I_{+}$ image was magnified above the threshold. This is shown in Table~\ref{lensing-2}. From Eq.(\ref{y_condition}) and Eq.(\ref{cross_section}) one can see that elementary cross sections for $I_{+}$ image magnified above threshold are higher than in the case of $I_{-}$ image. Therefore the numbers reported in Table~\ref{lensing-2} are much higher than analogous numbers in Table~\ref{lensing-1}. However, they should be compared with yearly detection rates of DCO inspiral events predicted for the ET (Table 1 and Table 2 of \citet{JCAP_ET2}). Such comparison shows that the magnification bias at the level of the full inspiral event catalog would be $0.001$. Therefore this would not affect much cosmological inferences drawn from such catalog. Figure~\ref{fig3} illustrates this effect by plotting together probability density of yearly detection rate of non-lensed (dashed line) sources and total prediction --- enriched by systems with $I_{+}$ image magnified (solid line). In particular, one can see how the lensing effect extends the high redshift tails of these distributions for all DCO systems except the BH-BH ones. However, the logarithmic scale was adopted in order to visualize this effect and the above mentioned claim of negligible magnification bias remains valid.

\section{Conclusions} \label{sec:conclusions}

We all hope that the new era of gravitational astronomy will soon be opened by the second generation of (upgraded) interferometric gravitational wave detectors. The expected benefits of direct detection of GW from astrophysical sources --- inspiralling DCO systems in the first place --- are much more than just seeing GW in flesh. They will provide unique tests for fundamental physics (e.g. alternative theories of gravity) and invaluable complementary tests of relativistic astrophysics (stellar and supermassive black holes) and cosmology. All the above mentioned is even more true for the ET which is expected to provide an extremely rich catalog of DCO inspirals detecting up to thousands of them per year. This way, ET will be sensitive to a population of sources at very high redshifts, allowing to study cosmological evolution of sources, the history of star formation and its dependence on the matter content of the Universe, and the development of large-scale structure in the Universe.

In this paper we extended our previous studies \citep{JCAP_ET1, JCAP_ET2} on gravitational lensing of such DCO inspiral events.
In particular we explored the case when DCO systems are intrinsically faint, i.e. having SNR parameter $\rho < 8$. This means that usually they would not be detected by ET, but the presence of gravitational lenses along the line of sight changes this situation and they could become visible to ET being magnified by lens. Our main result is that tens to hundreds of such extra events could be detected per year. Especially NS-NS (but also NS-BH) systems are affected by this mechanism and the ET's distance reach for signals from such coalescences broadens from $z \approx 1$ in the non-lensed case to $z \approx 4$ in the lensed case. This opens possibility to study star formation history covering substantially earlier epochs of the Universe. Lensed DCO events could also give us a unique information about masses and NS equation of state at much higher redshifts than one would be able to test without magnification by lenses. This inference would be based on individual systems not on statistical reasoning, so any single faint DCO inspiral would be useful. However, equally important will be to determine the Hubble constant or cosmic equation of state using catalogs of inspiral events. Therefore, one may worry about the adverse effect of contaminating the catalogs by lensed events (so called magnification bias). Hopefully our results show that the magnification bias is negligible and should not affect much cosmological inferences.

\section*{Acknowledgments}
The authors are grateful to the referee for very useful comments which allowed to improve the paper.
M.B. obtained approval of foreign talent introducing project in
China and gained special fund support of foreign knowledge
introducing project. He also gratefully acknowledges hospitality of
Beijing Normal University.
M.B. is partly supported by the Poland-China Scientific \& Technological Cooperation Committee Project No. 35-4.
M.B. also expresses his gratitude to the Kavli
Institute for Theoretical Physics China for hospitality, since this work
was substantially developed while M.B. was staying there.

Z.-H.Z. is supported by the Ministry of Science and Technology National Basic Science Program (Project 973) under Grants Nos. 2012CB821804 and 2014CB845806, the National Natural Science Foundation of China under Grants Nos. 11373014 and 11073005, and the Fundamental Research Funds for the Central Universities and Scientific Research Foundation of Beijing Normal University.


\end{document}